\documentclass[11pt,prd,aps,epsfig,floats,preprint,eqsecnum,nofootinbib]{revtex4-1}
\pdfoutput=1

\usepackage{color}
\usepackage[squaren]{SIunits}
\usepackage{amsmath}
\usepackage{amsfonts}
\usepackage{slashed}
\usepackage{graphicx}
\usepackage{subfigure}
\usepackage{rotating}

\usepackage{hyperref}
\usepackage{graphicx,amsmath,amssymb,epsfig,verbatim,mathrsfs,array,layout,textcomp,latexsym}
\usepackage{multirow}

\allowdisplaybreaks[1]

\renewcommand{\(}{\left(}
\renewcommand{\)}{\right)}
\renewcommand{\{}{\left\lbrace}
\renewcommand{\}}{\right\rbrace}
\renewcommand{\[}{\left\lbrack}
\renewcommand{\]}{\right\rbrack}

\newcommand{\refeq}[1]{Eq.~(\ref{eq:#1})}

\newcommand{\reffig}[1]{Fig.~\ref{fig:#1}}
\newcommand{\refsec}[1]{Section~\ref{sec:#1}}
\newcommand{\reftab}[1]{Table~\ref{tab:#1}}

\newcommand{\gfermi}{G_\mathrm{F}}
\newcommand{\GeV}{\,\mathrm{GeV}}
\newcommand{\MeV}{\,\mathrm{MeV}}

\newcommand{\wilson}[2][{}]{\mathcal{C}_{#2}^{\mathrm{#1}}}
\newcommand{\bra}[1]{\left\langle{#1}\right\vert}
\newcommand{\ket}[1]{\left\vert{#1}\right\rangle}

\begin{document}

\setlength{\parindent}{0pt}

\vspace{-0.2cm}
\hbox{DO-TH 11/23}
\hbox{EOS-2011-03}

\title{The Decay $\bar{B}\to\bar{K}\ell^+\ell^-$ at Low Hadronic Recoil \\and Model-Independent $\Delta B = 1$ Constraints}

\author{Christoph Bobeth} \affiliation{ Institute for Advanced Study \&
  Excellence Cluster Universe, Technische Universit\"at M\"unchen, D-85748
  Garching, Germany} \author{Gudrun Hiller} \author{Danny van Dyk}
\author{Christian Wacker} \affiliation{Institut f\"ur Physik, Technische
  Universit\"at Dortmund, D-44221 Dortmund, Germany}

\begin{abstract}
  We study the decay $\bar{B} \to \bar{K} \ell^+ \ell^-$  for $\ell = e, \mu,\tau$ with
   a softly recoiling kaon, that is, for high dilepton invariant masses
  $\sqrt{q^2}$ of the order of the $b$-quark mass. 
  This kinematic region can be treated within an operator product
  expansion and simplified using heavy quark symmetry, leading to systematic predictions for heavy-to-light processes such as $\bar{B} \to \bar{K}^{(*)} \ell^+ \ell^-$.
  We   show that the decay rates  of both $\bar{B} \to \bar{K}^{*} \ell^+ \ell^-$ and $\bar{B} \to \bar{K} \ell^+ \ell^-$ decays into light leptons depend on a common combination of short-distance coefficients. The
  corresponding CP-asymmetries are hence  identical. 
 Furthermore we present  low recoil predictions for $\bar{B} \to \bar{K} \ell^+ \ell^-$ 
 observables, including the flat term in the angular distribution which becomes sizable for taus. 
 We  work out model-independently the
   constraints on $\Delta B=1$ operators using the most recent data from the experiments BaBar, Belle, CDF and LHCb.  For constructive interference with the standard model, generic new physics is pushed up to scales above $44$ TeV at 95\% CL.
   Assuming none or small CP-violation we obtain a lower bound on the position of the zero of the forward-backward asymmetry of $\bar{B}^0 \to \bar{K}^{*0}
  \ell^+ \ell^-$ decays as $q_0^2 > 1.7\GeV^2$, which improves to $q_0^2 > 2.6 \GeV^2$
 for a standard model-like sign $b \to s \gamma$ amplitude.
  \end{abstract}

\maketitle

%
%
%--------+---------+---------+---------+---------+---------+---------+---------+
\section{Introduction}

Rare $B$-decays into dileptons  
are precision probes of  the standard model (SM) and the flavor sector and provide constraints on 
physics beyond the standard model (BSM). Important semileptonic modes in terms of experimental accessibility and theory control are  those into a $K$ or a $K^*$. The latter decays exhibit a rich phenomenology especially through angular analysis of subsequent decays $\bar{B} \to \bar{K}^* (\to \bar{K} \pi) \ell^+\ell^-$, see \cite{Bobeth:2011qn} for a recent summary. 
Decays into a pseudo-scalar meson $\bar{B} \to \bar{K} \ell^+\ell^-$, $\ell=e,\mu,\tau$ 
allow to perform a number of complementary measurements as well. Observables  include
the decay rate $\Gamma_\ell$, its CP-asymmetry $A_{\rm CP}^\ell$, the forward-backward asymmetry $A_{\rm FB}^\ell$ and  the flat term $F_H^\ell$ appearing in the 
angular distribution
\cite{Aubert:2006vb,Bobeth:2007dw,Ali:1999mm}
\begin{equation}
  \label{eq:angdist}
  \frac{1}{\Gamma_\ell} \frac{d\Gamma_\ell }{d\!\cos\theta_\ell}
    = \frac{3}{4} (1 - F_H^\ell) (1 - \cos^2\theta_\ell) 
    + \frac{1}{2} F_H^\ell + A_{\rm FB}^\ell \cos\theta_\ell \,  ,
\end{equation} 
or distributions in the dilepton mass thereof. (The angle $\theta_\ell$ is defined in \refsec{obsandang}.)

Previous systematic analyses of $\bar{B} \to \bar{K} \ell^+\ell^-$ distributions \cite{Bobeth:2007dw} focused on the region of large hadronic recoil, where QCD factorization (QCDF) applies \cite{Beneke:2001at, Beneke:2004dp}. The intermediate recoil region where
charmonium-resonances dominate the dilepton spectrum through
$\bar{B} \to  \bar{K}  (J/\Psi, \Psi^\prime)  \to \bar{K} \ell^+\ell^-$ decays has been studied
recently \cite{Khodjamirian:2010vf}. In this work we provide
a systematic analysis for the region of low hadronic recoil, that is, for large dilepton masses
$\sqrt{q^2}$ of the order of the $b$-quark mass, above the $\Psi^\prime$-peak.

The study of heavy-to-light decays at low recoil by means of a  local operator product expansion (OPE)  in $1/\sqrt{q^2}$ has been put forward by 
\cite{Grinstein:2004vb}, and recently \cite{Beylich:2011aq}, see also \cite{Buchalla:1998mt} for earlier mention on inclusive decays. The simultaneous matching onto heavy quark effective theory (HQET) and use of the improved Isgur-Wise form factor
relations \cite{Grinstein:2002cz} has been shown to be of benefit for  $\bar{B} \to \bar{K}^* \ell^+\ell^-$ decays
by allowing for the  design of specific low recoil observables with sensitivity
to either short- or long-distance physics as well as checks of the theory framework  \cite{Bobeth:2010wg,Bobeth:2011gi}.
Equally important, the theoretical accessibility of this kinematic region is
necessary for a full exploitation of the available and future rare $B$-decay data.
Note that roughly a similar amount of $\bar{B} \to \bar{K}^* \ell^+\ell^-$  data exists presently for the low and the large recoil region by all contemporary $B$-physics experiments, BaBar, Belle, CDF and LHCb.
The latter will soon be competitive and eventually take over in statistics
as indicated by the recent preliminary results on the
angular distribution of $\bar{B}^0\to\bar{K}^{*0} \ell^+\ell^-$ decays \cite{LHCb:2011,Blake:2011ii}.
As for $\bar{B} \to \bar{K} \ell^+\ell^-$ decays, LHCb 
reported  so far $35\pm 7$ events of $B^+\to K^+\mu^+\mu^-$ within a recorded luminosity of $37{\rm pb}^{-1}$ during the 2010 LHC run \cite{Golutvin:2011}.

The plan of this paper is to work out and explore the phenomenology of $\bar{B} \to
\bar{K} \ell^+\ell^-$ decays at low hadronic recoil. We give in
\refsec{at-low-recoil} the heavy-quark form factor relation, collect the expressions for the
decay amplitude at low recoil and discuss the observables relevant to our
model-independent framework. In \refsec{numerical-evaluation} we present SM
predictions including a discussion of their uncertainties.
In \refsec{model-independent-analysis} we derive constraints on the $\Delta B=1$ short-distance couplings entering $\bar{B} \to \bar{K}^{(*)}\ell^+\ell^-$ decays using the most recent
experimental low and large recoil data. We conclude in \refsec{conclusion}.
Details on the form factors are given in Appendix \ref{app:FFs}.

%
%
%--------+---------+---------+---------+---------+---------+---------+---------+
\section{The Decay $\bar{B}\to\bar{K}\ell^+\ell^-$ at Low Recoil}
\label{sec:at-low-recoil}

We use an effective $\Delta B =1$ Hamiltonian to describe the  flavor-changing $b \to s \ell^+ \ell^-$transitions as
\begin{align}
  \label{eq:Heff}
  \mathcal{H}_{\rm eff}
    & = -\frac{G_{\rm F}}{\sqrt{2}} \frac{\alpha_e}{\pi} V_{tb} V_{ts}^* \\
  \nonumber
    & \times
        \(\wilson{7} \frac{m_b}{e} \[\bar{s} \sigma_{\mu\nu} P_{R} b\] F^{\mu\nu}
        + \wilson{9} \[\bar{s} \gamma_\mu P_L b\]\[\bar{\ell}\gamma^\mu \ell\]
        + \wilson{10} \[\bar{s} \gamma_\mu P_L b\]\[\bar{\ell}\gamma^\mu\gamma_5 \ell\]
        + \rm{h.c.} \) + \dots\, .
\end{align}
Here, the ellipses denote tree-level induced or subdominant contributions which
we assume to be SM-like.  Furthermore, $V_{ij}$ denote CKM-elements and
$m_b$ the $\overline{\rm MS}$-mass of the $b$-quark. We allow the Wilson
coefficients $\wilson{7,9,10}$ to be complex-valued  to account for  CP-violation beyond the SM. We assume them throughout this work to be evaluated at the scale $\mu=m_b$.
For further details we refer to  previous low recoil works employing the same notation \cite{Bobeth:2010wg,Bobeth:2011gi}.

%
%--------+---------+---------+---------+---------+---------+---------+---------+
\subsection{The Improved Isgur-Wise Relation}

The matrix elements of $\bar{B} \to \bar{K}$ transitions can be parameterized in
terms of three $q^2$-dependent form factors $f_{+,T,0}$ which are defined in
Appendix \ref{app:FFs}. Following \cite{Grinstein:2002cz, Grinstein:2004vb}, the
QCD operator identity
\begin{align}
  \label{eq:opID}
  i \partial^\nu (\bar{s}\, i \sigma_{\mu \nu}\, b) & =
   i \partial_\mu (\bar{s} b)
   - m_b  (\bar{s}\, \gamma_\mu\, b)
   - 2 (\bar{s}\, i \!\stackrel{\leftarrow}{D}_{\mu} \! b)
\end{align}
allows to derive an improved Isgur-Wise relation between $f_T$ and $f_+$,
\begin{align}
  f_T(q^2, \mu) &
  = \frac{m_B\, (m_B + m_K)}{q^2} \[ \kappa(\mu) f_+(q^2) + \frac{2\, \delta_+^{(0)}(q^2)}{m_B} \]
    + \mathcal{O}\left(\alpha_s \frac{\Lambda}{m_b}, \frac{\Lambda^2}{m_b^2} \right) \label{eq:IW-relation-sl-terms}
\\[0.2cm]
  & = \frac{m_B\, (m_B + m_K)}{q^2}\,  \kappa(\mu)\, f_+(q^2)
    + \mathcal{O}\left( \frac{\Lambda}{m_b} \right) \, ,
  \label{eq:IW-relation}
\end{align}
in agreement  with \cite{HurthWyler2005}. (We neglect the mass of the strange quark.) 
Here, we denote by $m_B,m_K$ the $B$-meson and kaon mass, respectively, and neglect in the second line the subleading HQET form factor $\delta_+^{(0)}$, see Appendix \ref{app:FFs}.
The $1/q^2$-factor on the right-hand side of Eqs.~(\ref{eq:IW-relation-sl-terms}) and (\ref{eq:IW-relation}) is of kinematical origin, related to the definition of $f_T$, \refeq{fT}. 
The $\mu$-dependent coefficient $\kappa$ reads, including $\mathcal{O}(\alpha_s)$ corrections,
 as
\begin{align} \label{eq:kappa}
  \kappa(\mu) & = \(1 + 2\, \frac{D_0^{(v)}(\mu)}{C_0^{(v)}(\mu)}\)\frac{m_b(\mu)}{m_B}
\end{align}
with the HQET Wilson coefficients $C_0^{(v)}, D_0^{(v)}$ given in
\cite{Grinstein:2004vb}. At $\mu = m_b$ holds $\kappa = 1 + \mathcal{O}(\alpha_s^2)$.
While deriving Eq.~(\ref{eq:IW-relation}) 
one finds that the form factor $f_0$ is a power correction
\begin{align} \label{eq:f0}
f_0(q^2) = 0 + \mathcal{O}\left( \frac{\Lambda}{m_b} \right) ,
\end{align}
as expected from heavy quark symmetry \cite{Isgur:1990kf}.
%

%
%--------+---------+---------+---------+---------+---------+---------+---------+
\subsection{The $\bar{B}\to\bar{K}\ell^+\ell^-$ Matrix Element}

Beyond the contributions from the operators in \refeq{Heff}, the
amplitude of $\bar{B}\to\bar{K}\ell^+\ell^-$ also receives contributions from
current-current and QCD-penguin operators. As a result of the OPE in $1/\sqrt{q^2}$
 for the leading dimension 3-operators, these contributions can be
taken into account by effective Wilson coefficients $\wilson[eff]{7,9}$
\cite{Beylich:2011aq}, whereas sub-leading contributions enter at dimension 5
and are suppressed by $(\Lambda/m_b)^2 \sim 2\%$. Here we follow
\cite{Grinstein:2004vb}  and subsequent low recoil $\bar{B}\to \bar{K}^*\ell^+\ell^-$ works  \cite{Bobeth:2010wg,Bobeth:2011gi} and perform an additional matching onto HQET and use the
form factor relation \refeq{IW-relation}. 
The  effective coefficients   are then given as  \cite{Bobeth:2011gi}
\begin{align}
  \label{eq:c7effGP}
  \wilson[eff]{7} & =
  \wilson{7} -
  \frac{1}{3} \[ \wilson{3} + \frac{4}{3}\,\wilson{4} + 20\,\wilson{5}
    + \frac{80}{3}\wilson{6} \]
    + \frac{\alpha_s}{4 \pi} \[ \left(\wilson{1} - 6\,\wilson{2}\right) A(q^2)
    - \wilson{8} F_8^{(7)}(q^2)\]\,,\\
  \label{eq:c9effGP}
  \wilson[eff]{9} & =
    \wilson{9} +
     h(0, q^2) \[ \frac{4}{3}\, \wilson{1} + \wilson{2} + \frac{11}{2}\, \wilson{3}
    - \frac{2}{3}\, \wilson{4} + 52\, \wilson{5} - \frac{32}{3}\, \wilson{6}\]
\\
  & - \frac{1}{2}\, h(m_b, q^2) \[ 7\, \wilson{3} + \frac{4}{3}\, \wilson{4}
      + 76\, \wilson{5} + \frac{64}{3}\, \wilson{6} \]
    + \frac{4}{3} \[ \wilson{3} + \frac{16}{3}\, \wilson{5} + \frac{16}{9}\, \wilson{6} \]
\nonumber\\[0.2cm]
  & + \frac{\alpha_s}{4 \pi} \[ \wilson{1} \left(B(q^2) + 4\, C(q^2)\right)
    - 3\, \wilson{2}\left(2\, B(q^2) - C(q^2)\right) - \wilson{8}F_8^{(9)}(q^2) \]
\nonumber\\[0.2cm]
  & + 8\, {\frac{m_c^2}{q^2}  \[ \left( \frac{4}{9}\,\wilson{1}
    + \frac{1}{3}\,\wilson{2} \right)(1+\hat \lambda_u) + 2\,\wilson{3} + 20\,\wilson{5} \] } .
\nonumber
\end{align}
These include the NLO QCD matching corrections and doubly Cabibbo-suppressed
contributions proportional to 
$\lambda_u = V_{ub}^{} V_{us}^{*}/(V_{tb}^{} V_{ts}^{*})$. The latter  are
responsible for the tiny amount of CP-violation in the SM in $b\to s$ transitions.

The $\bar{B}\to \bar{K}\ell^+\ell^-$ decay amplitude
\cite{Bobeth:2007dw} simplifies within the SM operator basis \refeq{Heff} after applying
the form factor relation \refeq{IW-relation} to
\begin{align}
  \mathcal{A}(\bar{B}\to \bar{K}\ell^+\ell^-)
    & = i \frac{\gfermi \alpha_e}{\sqrt{2} \pi}\, V_{tb}^{} V^*_{ts}\, f_+(q^2)
        \big[
           F_V\, p^\mu (\bar{\ell} \gamma_\mu \ell)
         + F_A\, p^\mu (\bar{\ell} \gamma_\mu\gamma_5 \ell)
         + F_P\, (\bar{\ell} \gamma_5 \ell) \big]\,,
  \label{eq:hadr-ME}
\end{align}
where
\begin{align}
  F_A & = \wilson{10}\,, &
  F_V & = \wilson[eff]{9} + \kappa\, \frac{2\, m_b\, m_B}{q^2}\, \wilson[eff]{7}\,,
  \nonumber
\end{align}
\begin{align}
   F_P & = -m_\ell \[1 + \frac{m_B^2-m_K^2}{q^2}\(1-\frac{f_0}{f_+}\)\]\,
          \wilson{10}\, ,
\end{align}
and $p^\mu$ denotes the 4-momentum of the $B$-meson and $m_\ell$ the lepton mass.

The use of  \refeq{IW-relation}  introduces an
uncertainty of  $\mathcal{O}(\Lambda/m_b)$  in the term proportional to $\wilson[eff]{7}$ in
$F_V$, however, the phenomenological impact of this uncertainty is additionally suppressed by 
$\wilson[eff]{7}/\wilson[eff]{9}$, which is $\sim 0.1$ in the SM. Further
subleading $1/m_b$ contributions to the amplitude $F_V$ in \refeq{hadr-ME} itself
receive an additional suppression of $\alpha_s$ \cite{Grinstein:2004vb}. The associated uncertainties are included in our phenomenological analysis 
following the procedure described in \cite{Bobeth:2011gi}.

The coefficient $F_P$ is
suppressed by $m_\ell/m_B$ and hence small for light leptons. Since the form factor $f_0$
enters $F_P$ only, its impact for phenomenological implications is negligible except for
 taus.

%
%--------+---------+---------+---------+---------+---------+---------+---------+
\subsection{Observables and Angular Distribution \label{sec:obsandang}}

Continuing along the lines of \cite{Bobeth:2007dw}, we write the
differential $\bar{B}\to\bar{K}\ell^+\ell^-$  decay distributions as
\begin{align}
  \frac{{\rm d}^2 \Gamma_\ell [\bar{B}\to\bar{K}\ell^+\ell^-]}
       {{\rm d} q^2\,{\rm d} \cos\theta_\ell}
    & = a_\ell(q^2) + b_\ell(q^2) \cos\theta_\ell + c_\ell(q^2) \cos^2\theta_\ell\,,
\\[0.2cm]
  \frac{{\rm d} \Gamma_\ell [\bar{B}\to\bar{K}\ell^+\ell^-]}{{\rm d} q^2}
    & = 2\[a_\ell(q^2) + \frac{1}{3} c_\ell(q^2) \]\,
\end{align}
with $q^2$-dependent observables $a_\ell$, $b_\ell$ and $c_\ell$.
The angle $\theta_\ell$ is defined as the angle between the $\bar{B}$-direction
and the $\ell^-$-direction in the $\ell^+\ell^-$ rest frame. Within the SM
operator basis \refeq{Heff}
\begin{align}
  \frac{a_\ell}{\Gamma_0 \sqrt{\lambda} \beta_\ell f_+^2} & =
    \frac{\lambda}{4} \(|F_A|^2 + |F_V|^2\)
\nonumber\\
    & \quad + 2 m_\ell (m_B^2 - m_K^2 + q^2) {\rm Re}(F_P^{} F_A^*)
    + 4\, m_\ell^2\, m_B^2 |F_A|^2
    + q^2 |F_P|^2\,,
    \nonumber\\[0.2cm]
    b_\ell & =0 ,
\nonumber\\[0.2cm]
  \frac{c_\ell}{\Gamma_0 \sqrt{\lambda} \beta_\ell f_+^2} & =
    - \beta_\ell^2\, \frac{\lambda}{4} \(|F_A|^2 + |F_V|^2\)
\end{align}
with
\begin{align}
  \Gamma_0 &
    = \frac{G_{\rm F}^2 \alpha_e^2 |V_{tb}^{} V^*_{ts}|^2}{2^9 \pi^5 m_B^3}\,, &
  \beta_\ell &
    = \sqrt{1 - \frac{4 m_\ell^2}{q^2}}\,,
\end{align}
\begin{align}
  \lambda \equiv \lambda(m_B^2, m_K^2, q^2)
     & = m_B^4 + m_K^4 + q^4 - 2\(m_B^2 m_K^2 + m_B^2 q^2 + m_K^2 q^2\)\,.
  \nonumber
\end{align}
The terms in the  $\bar{B}\to \bar{K}\ell^+\ell^-$ angular distribution \refeq{angdist} can then be obtained as \cite{Bobeth:2007dw}
\begin{align}
\Gamma_\ell & =2 \int_{q^2_{\rm min}}^{q^2_{\rm max}} dq^2\ (a_\ell + \frac{1}{3} c_\ell ) ,\\
A_{\rm FB}^\ell & =0 , \label{eq:AFB}\\
F_H^\ell & = \frac{2}{\Gamma_\ell}   \int_{q^2_{\rm min}}^{q^2_{\rm max}} dq^2 ( a_\ell +  c_\ell ) .
\label{eq:flat}
\end{align}
The forward-backward asymmetry of $\bar{B}\to \bar{K}\ell^+\ell^-$ decays is zero
within the SM operator basis\footnote{Non-zero values in the SM originate from
  QED corrections and are tiny \cite{Bobeth:2007dw,Demir:2002cj}.} and 
  $F_H^\ell$ is suppressed by lepton mass, see below. However both observables receive contributions  from  scalar and tensor operators, and can signal such BSM effects  \cite{Bobeth:2007dw,Alok:2008wp}.

For  $\ell = e,\mu$ we can safely neglect the lepton mass. In this limit $\beta_\ell=1$ and 
\begin{align}
   a_\ell & = \Gamma_0 \frac{\sqrt{\lambda}^3}{4} f_+^2\, \rho_1\,, &
   c_\ell & = -a_\ell\,,
\end{align}
with the short-distance coefficient
\begin{align}
  \rho_1 &
   = \left|\wilson[eff]{9} + \kappa \frac{2\, m_b\, m_B}{q^2}\,\wilson[eff]{7}\right|^2
      + \left|\wilson{10}\right|^2 .
\end{align}
The differential decay rate at low recoil can then be written as
\begin{align} \label{eq:q2diff0}
  \frac{{\rm d} \Gamma_\ell [\bar{B}\to\bar{K}\ell^+\ell^-]}{{\rm d} q^2}
    & =  \Gamma_0 \frac{\sqrt{\lambda}^3}{3} f_+^2\, \rho_1\, ,
    \end{align}
    and $F_H^\ell$ vanishes for  $m_\ell=0$.
    
The SM value of $\rho_1$ and its associated uncertainties  have been presented earlier
 in the discussion of $\bar{B}\to\bar{K}^*\ell^+\ell^-$ decays at
low recoil \cite{Bobeth:2010wg}. The appearance of the same short-distance factor $\rho_1$ in both vector and pseudoscalar final states  uniquely correlates the
decays $\bar{B}\to\bar{K}\ell^+\ell^-$ and
$\bar{B}\to\bar{K}^*\ell^+\ell^-$ into light leptons. As a consequence, the
CP-asymmetry of the decay rate, $A_{\rm CP}^\ell$, is identical 
\begin{align} \label{eq:acp1}
  A_{\rm CP}^\ell[\bar{B}\to\bar{K}\ell^+\ell^-]
  & = \frac{{\rm d} \Gamma_\ell[\bar{B}\to\bar{K}\ell^+\ell^-]/{\rm d} q^2 - {\rm d} {\Gamma}_\ell [{B}\to {K}\ell^+\ell^-]/{\rm d} q^2}
           {{\rm d} \Gamma_\ell[\bar{B}\to\bar{K}\ell^+\ell^-]/{\rm d} q^2 + {\rm d} {\Gamma}_\ell[{B}\to{K}\ell^+\ell^-]/{\rm d} q^2}
\\[0.2cm]
  & = \frac{\rho_1 - \bar{\rho}_1}{\rho_1 + \bar{\rho}_1}
    = a_{\rm CP}^{(1)}[\bar{B}\to\bar{K}^*\ell^+\ell^-] ,
  \nonumber
\end{align}
where $\bar{\rho}_1$ is obtained from $\rho_1$ by complex conjugation of the
weak phases, {\it i.e.}, the CKM matrix elements and the Wilson coefficients
$\wilson{i}$.  The CP-asymmetry $a_{\rm CP}^{(1)}$ is
form factor-free and 
essentially vanishes in the SM $|a_{\rm CP}^{(1)}|_{\rm SM} \lesssim 10^{-4}$  \cite{Bobeth:2011gi}.

Keeping the lepton mass finite we obtain
\begin{align}
  a_\ell & = \frac{\Gamma_0}{4} \sqrt{\lambda}^3\, \beta_\ell f_+^2
  \[\rho_1 - \frac{4\, m_\ell^2}{q^2} |\wilson{10}|^2 {\cal F}_0 \]\,,
&%\nonumber\\[0.2cm]
  c_\ell & = -\frac{\Gamma_0}{4}\sqrt{\lambda}^3\, \beta_\ell^3 f_+^2\, \rho_1\,,
\end{align}
with
\begin{align}
  {\cal F}_0 & = 1 - \frac{(m_B^2 - m_K^2)^2}{\lambda}  \( \frac{f_0}{f_+} \)^2\,.
\end{align}
The second term  in ${\cal{F}}_0$ is order one despite $f_0$ being a power correction, see \refeq{f0},
because of  the phase space enhancement by the factor $\lambda \sim {\cal{O}}( \Lambda^2 m_B^2)$ in the denominator.

The  $q^2$-differential flat term of the angular distribution 
\begin{align} \label{eq:flatq2}
  F_H^\ell(q^2) &
   = \frac{a_\ell + c_\ell}{a_\ell + \frac{1}{3}c_\ell}
   = \frac{6\, m_\ell^2}{q^2} \times
     \frac{\rho_1 - |\wilson{10}|^2 {\cal F}_0}
          {\rho_1 + \frac{2\, m_\ell^2}{q^2} \left(\rho_1 - 3\, |\wilson{10}|^2 {\cal F}_0 \right)}
\end{align}
can be sizable for $\ell=\tau$, as we show explicitly in \refsec{numerical-evaluation} for the SM.
The $F_H^\tau(q^2)$-observable is  complementary
to the $\bar{B}\to \bar{K}\ell^+\ell^-$ branching ratios. From combined analysis with 
the branching ratios of $\bar{B}\to \bar{K}^{(*)} \ell^+\ell^-$ for $\ell=e,\mu$  decays the coefficient
$\rho_1$ could be extracted together with the magnitude of $\wilson{10}$, see also  \cite{Skands:2000ru}, given sufficient control of $f_0/f_+$.

%
%--------+---------+---------+---------+---------+---------+---------+---------+
\section{Phenomenological analysis}

For the phenomenological analysis of the various rare $B$ decay observables we use the
numerical inputs given in Table \ref{tab:numericConstants}. All numerical results
presented  are obtained with the EOS flavor tool \cite{EOS}.
All experimental $\bar{B} \to
\bar{K}^{(*)} \ell^+\ell^-$ results shown and used are CP-averages, and should silently be understood as those.

%--------+---------+---------+---------+---------+---------+---------+---------+
\subsection{Standard Model Predictions for $\bar{B} \to
\bar{K} \ell^+\ell^-$}
\label{sec:numerical-evaluation}

\begin{table}
\begin{tabular}{|llr|llr|}
\hline
$A$                      & $0.812^{+0.013}_{-0.027}$       &
\cite{Charles:2004jd}   &
$\lambda$                & $0.22543\pm0.00077$             &
\cite{Charles:2004jd}   \\
$\bar{\rho}$             & $0.144\pm{0.025}$               &
\cite{Charles:2004jd}   &
$\bar{\eta}$             & $0.342^{+0.016}_{-0.015}$       &
\cite{Charles:2004jd}   \\
$\alpha_s(M_Z)$          & $0.11762$                       & &
$\tau_{B^+}$             & $1.638~\pico\second$            &
\cite{Nakamura:2010zzi} \\
$\alpha_e(m_b)$          & $1/133$                         & &
$\tau_{B^0}$             & $1.525~\pico\second$            &
\cite{Nakamura:2010zzi} \\
$m_c(m_c)$               & $(1.27^{+0.07}_{-0.09})~\GeV$   &
\cite{Nakamura:2010zzi}   &
$m_{B^+}$                & $5.2792~\GeV$                   &
\cite{Nakamura:2010zzi} \\
$m_b(m_b)$               & $(4.19^{+0.18}_{-0.06})~\GeV$   &
\cite{Nakamura:2010zzi}   &
$m_{B^0}$                & $5.2795~\GeV$                   &
\cite{Nakamura:2010zzi} \\
$m_t^{\rm pole}$         & $(173.3\pm1.3)~\GeV$            &
\cite{:2009ec}          &
$m_{K^+}$                & $0.494~\GeV$                    &
\cite{Nakamura:2010zzi} \\
$m_e$                    & $0.511~\MeV$                    &
\cite{Nakamura:2010zzi} &
$m_{K^0}$                & $0.498~\GeV$                    &
\cite{Nakamura:2010zzi} \\
$m_\mu$                  & $0.106~\GeV$                    &
\cite{Nakamura:2010zzi} &
$M_W$                    & $(80.399\pm0.023)~\GeV$         &
\cite{Nakamura:2010zzi} \\
$m_\tau$                 & $1.777~\GeV$                    &
\cite{Nakamura:2010zzi}   &
$\sin^2\theta_W$         & $0.23116\pm0.00013$             &
\cite{Nakamura:2010zzi} \\
\hline
\end{tabular}
\caption{The numerical input used in our analysis. We neglect the mass of the
  strange quark. The superscript $^0(^+)$ refers to neutral (charged) parameters.
 \label{tab:numericConstants}
}
\end{table}

The SM predictions for the $\bar{B} \to
\bar{K} \ell^+\ell^-$  observables at low recoil in the framework  described in \refsec{at-low-recoil} are given in Table \ref{tab:Predictions}. The integration region is
$14.18\,\GeV^2 \leq q^2 \leq (m_B - m_K)^2$, {\it i.e.}, 
above the $\Psi^\prime$-resonance. 
We employ
the form factors $f_+$ and $f_0$ -- the latter is required for decays to massive leptons only -- from light cone QCD sum rule (LCSR) calculations and
a $q^2$-shape by a series expansion from \cite{Khodjamirian:2010vf} extrapolated to high $q^2$.
Our choice is motivated by the good agreement with preliminary unquenched lattice results \cite{Liu:2011ra} and the availability of uncertainties for the parameterization of the form factors, see Appendix \ref{app:FFs} for details. The dependence on the tensor form factor has been removed by means of \refeq{IW-relation}.

The uncertainties given in Table \ref{tab:Predictions} originate from the form factors (FF), specifically the variation of the inputs \reftab{ffinput} to the parameterization \refeq{ffparam},  the CKM matrix elements (CKM), and the short-distance parameters $m_t^{\rm pole}$, $M_W$, $\sin^2 \theta_W$ as well as the scale $\mu$ varied within
$[\mu_b/2, 2\mu_b] $ with $\mu_b = 4.2\, \GeV$ collectively denoted as  (SD). We evaluate the SM Wilson coefficients at next-to-next-to-leading order. In addition, we take into account the (unknown) subleading $1/m_b$-corrections to the $\bar{B} \to \bar{K} \ell^+\ell^-$ matrix elements and form factor relation \refeq{IW-relation} (SL).  The latter uncertainty is estimated with the 
procedure outlined in the
Appendix of \cite{Bobeth:2011gi} adopted to $\bar{B}\to \bar{K}\ell^+\ell^-$ decays.
The residual CKM uncertainty from $\hat \lambda_u$ in
$F_H^\ell$  has been dropped  in Table \ref{tab:Predictions} because it is below the given precision. To the accuracy given  in Table \ref{tab:Predictions} SM CP-violation is too small to induce any effect.

At high $q^2$ weak annihilation topologies
are strongly suppressed by $(\Lambda/m_b)^3$ \cite{Beylich:2011aq}, and thus isospin breaking effects are dominated by the differences in the lifetime.
The branching ratios of decays of neutral $B$-mesons are obtained by rescaling the corresponding branching ratios  of charged $B$ decays  in Table \ref{tab:Predictions}  with
the ratio of lifetimes $\tau_{B^0}/\tau_{B^+}$.  In the numerics  we keep  as well the very small effect of isospin breaking by the  masses $m_B$ and $m_K$. 
\begin{table}
\begin{tabular}{|c|c|c|c|c|c|}
\hline
Observable & Central Value & FF & SD & SL & CKM \tabularnewline
\hline
\hline
$10^{7}\times \mathcal{B}(B^-\to K^-\ell^+\ell^-)$                  & $1.04$ & $_{-0.27}^{+0.60}$ & $_{-0.02}^{+0.04}$ & $_{-0.03}^{+0.03}$ & $_{-0.07}^{+0.04}$\tabularnewline
$10^{7}\times \mathcal{B}(B^-\to K^-\tau^+\tau^-) $            & $1.26$ & $_{-0.21}^{+0.40}$ & $_{-0.03}^{+0.04}$ & $_{-0.02}^{+0.02}$ & $_{-0.09}^{+0.05}$\tabularnewline
$10^{3}\times  F_{\rm H}^\mu  $                                 & $7.50$ & $_{-2.61}^{+2.85}$ & $_{-0.10}^{+0.04}$ & $_{-0.10}^{+0.10}$ & -                 \tabularnewline
$10^{1}\times  F_{\rm H}^\tau  $                                & $8.90$ & $_{-0.45}^{+0.33}$ & $_{-0.02}^{+0.01}$ & $_{-0.02}^{+0.02}$ & -                 \tabularnewline
\hline
\end{tabular}
\caption{The SM branching ratios 
  and  flat terms $ F_{H}^{\ell}$ \refeq{flat}  integrated 
  from $14.18\,\GeV^2$ to $(m_B - m_K)^2$. For a description of the uncertainties see text. The branching ratios for  the light leptons $\ell=e$ and $\ell=\mu$
  are equal within the given precision (finite lepton masses are taken into account). 
  The branching ratios for neutral $B$ decays are obtained by rescaling the corresponding charged ones with
  $\tau_{B^0}/\tau_{B^+}$. All branching ratios are identical within the given accuracy to the ones of the  corresponding CP-conjugated decays.
  \label{tab:Predictions}
}
\end{table}

\begin{figure}[t]
  \includegraphics[width=.9\textwidth]{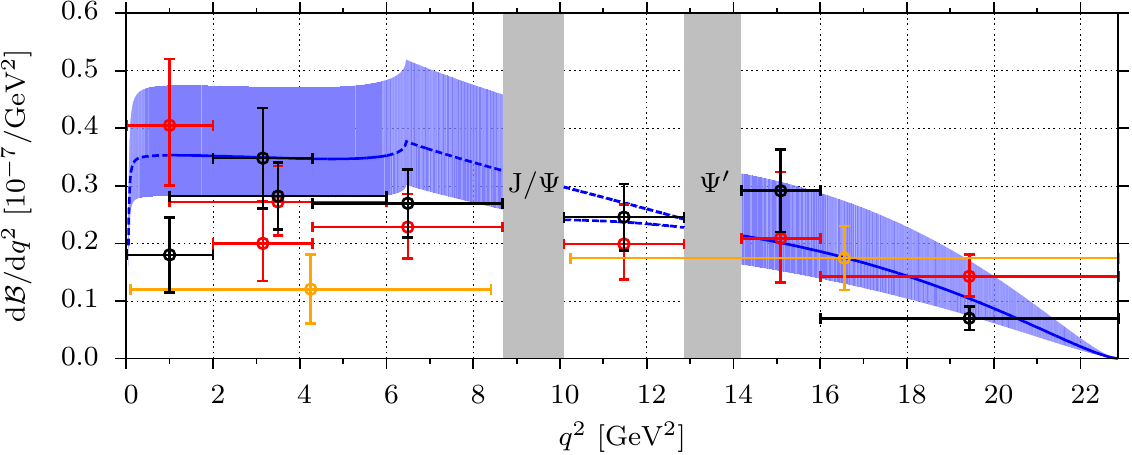}
  \caption{The SM branching ratio ${\rm d}\mathcal{B}(B^-\to K^-\mu^+\mu^-)/{\rm d}q^2$  using form factors from \cite{Khodjamirian:2010vf}
    (blue solid and dashed lines) alongside the measurements by Belle 
  \cite{:2009zv}
    (red points), CDF \cite{Aaltonen:2011qs} (black points) and BaBar \cite{Aubert:2006vb}
    (orange points), see footnote \ref{admixure}. The blue band shows the theoretical uncertainties described in the text added
    in quadrature. 
        The vertical (grey) bands are
  the experimental veto regions \cite{:2009zv,Aaltonen:2011qs} to remove contributions
  from $\bar B \to J/\Psi (\to \mu^+ \mu^-) \bar K$ (left-hand band) and $\bar B
  \to \Psi^\prime (\to \mu^+ \mu^-) \bar K$ (right-hand band). 
  The dashed (blue) lines below the $J/\Psi$  (between the
  charmonium bands) correspond to theory extrapolations from the low recoil (low and large recoil) region.
   \label{fig:diff-br}}
\end{figure}

\begin{figure}[t]
  \includegraphics[width=.9\textwidth]{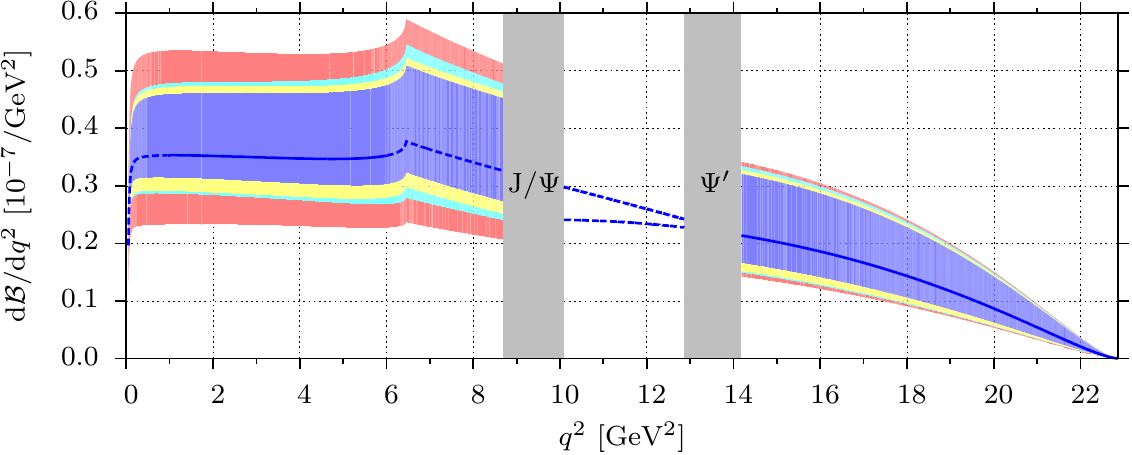}
  \caption{The SM branching ratio  ${\rm d}\mathcal{B}(B^-\to K^-\mu^+\mu^-)/{\rm d}q^2$
  with the linearly added uncertainties from the form factors (blue), the CKM matrix elements (yellow), the short-distance input (cyan) and the subleading $1/m_b$ corrections (red), see also \reffig{diff-br}.
  \label{fig:diff-br-uncert}}
\end{figure}

\begin{figure}[t]
  \includegraphics[width=.9\textwidth]{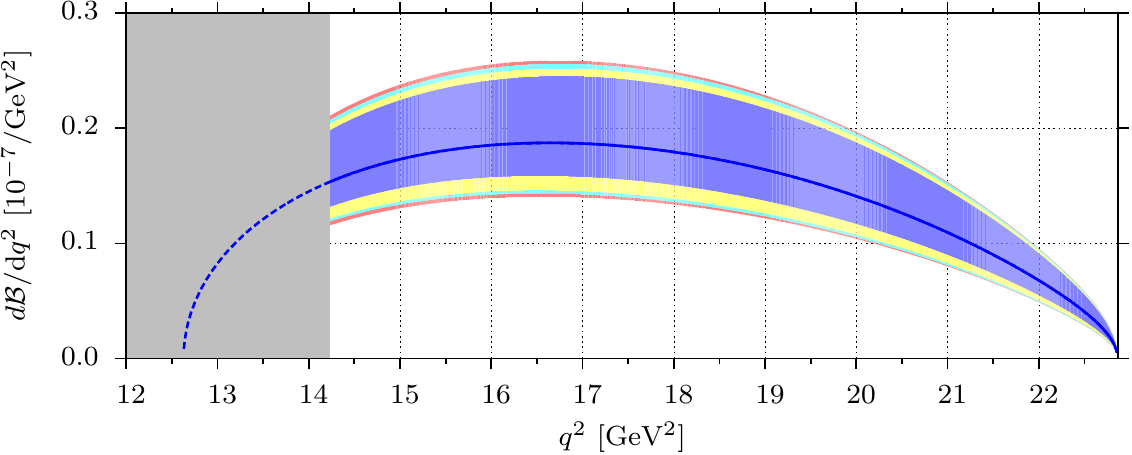}
  \caption{The SM branching ratio  ${\rm d}\mathcal{B}(B^-\to K^-\tau^+\tau^-)/{\rm d}q^2$
  with  linearly added uncertainties,   see \reffig{diff-br-uncert}.
  The grey band refers to the cut from BaBar's search for  this channel \cite{Walsh:2011}.
    \label{fig:diff-br-uncert-tau}}
\end{figure}

\begin{figure}[t]
  \includegraphics[width=.9\textwidth]{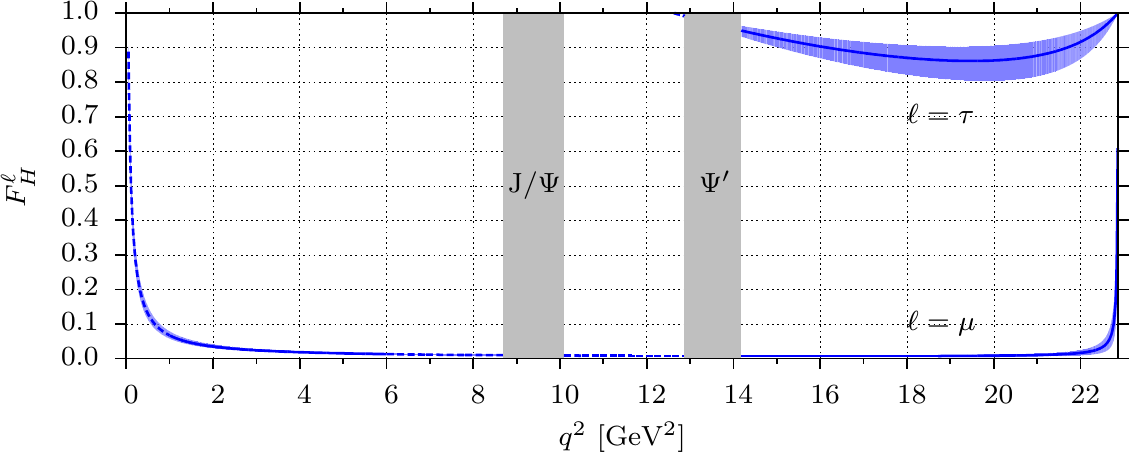}
  \caption{The flat term $F_H^\ell(q^2)$  for $B^-\to K^-\ell^+\ell^-$ decays  given in \refeq{flatq2}
  for $\ell=\mu$ (lower curve) and
    $\ell=\tau$ (upper band) in the SM. The theoretical uncertainties are added in quadrature.
    \label{fig:flat_term}}
\end{figure}

We show the SM
predictions for ${\rm d}\mathcal{B}(B^-\to K^-\mu^+\mu^-)/{\rm d}q^2$ in \reffig{diff-br} versus existing data\footnote{ \label{admixure}The data on the binned $\bar{B} \to
\bar{K} \ell^+\ell^-$ branching ratios by Belle and BaBar are publicly available for an unknown admixture of charged and neutral $B$ decays only. We find
differences of the order of the  scan resolution in the constraints Figs.~\ref{fig:constraints-c9-c10} 
 and \ref{fig:constraints-c9-c10-history} when interpreting the Belle data as either purely charged or  as purely neutral $B$ decays.
This effect will be even less important in the future with improved statistics.}
by Belle   \cite{:2009zv} (red points),  CDF \cite{Aaltonen:2011qs} (black points) and BaBar \cite{Aubert:2006vb}
    (orange points)
and with individual
uncertainty budgets in \reffig{diff-br-uncert}. In \reffig{diff-br-uncert-tau} we show
the differential SM branching ratio into ditaus.
The evaluation of observables in the large recoil region is carried out within the framework
of QCDF \cite{Beneke:2001at,Beneke:2004dp}. We employ the same $\bar{B} \to
\bar{K} \ell^+\ell^-$ form factors used at low recoil, {\it i.e.,} the ones from Ref.~\cite{Khodjamirian:2010vf}.
The treatment of the subleading corrections  for low $q^2$ is adapted from \cite{Egede:2008uy}.
 The flat term $F_H^\ell(q^2)$ for both muons and taus is shown in \reffig{flat_term}.
The growth of $F_H^\mu(q^2)$  at both ends of the spectrum results from a finite muon mass.
While at low $q^2$ such effects are kinematically enhanced by 
$m_\mu^2/q^2$, the effect at the very
high $q^2$-end  is induced by the vanishing decay rate in the denominator.
For electrons  these effects are further suppressed by $m_e^2/m_\mu^2 \sim 10^{-5}$, and negligible.

%--------+---------+---------+---------+---------+---------+---------+---------+
\subsection{Model-Independent Analysis}
\label{sec:model-independent-analysis}

\begin{figure}[t]
  \includegraphics[width=.66\textwidth]{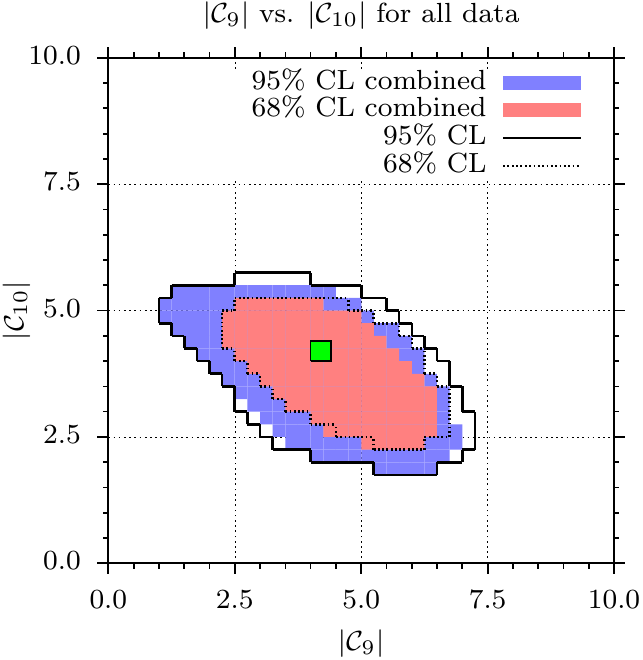}
\caption{Constraints on $|\wilson{9}|$ and $|\wilson{10}|$ 
    from  the combined analysis 
    of  $\bar{B}\to\{\bar{K},\bar{K}^{*},X_s\}\ell^+\ell^-$
    decays at $68\%$ CL (red area)  and $95\%$ CL (blue area). The $68\%$ CL (dotted) and $95\%$ CL (solid) contours  without using
    $\bar B\to \bar{K} \ell^+\ell^-$  decays are shown as well.
    The green square marks the SM prediction.
    \label{fig:constraints-c9-c10}
  }
\end{figure}

\begin{figure}[t]
  \includegraphics[width=.32\textwidth]{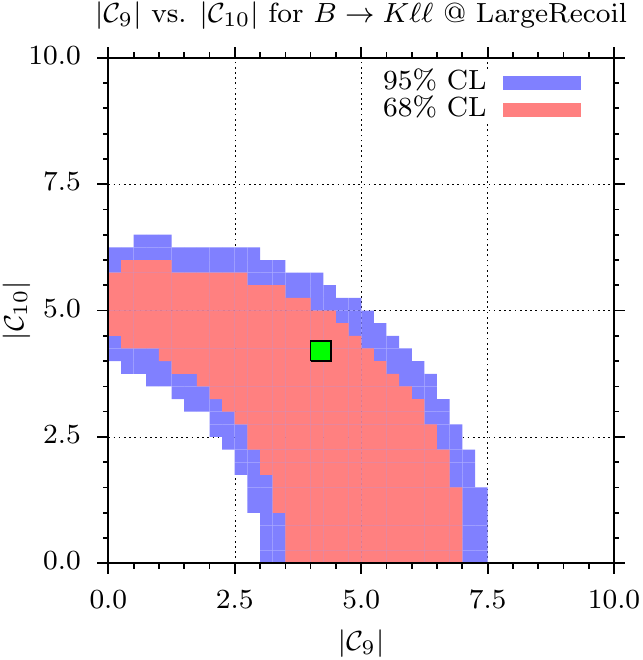}
  \includegraphics[width=.32\textwidth]{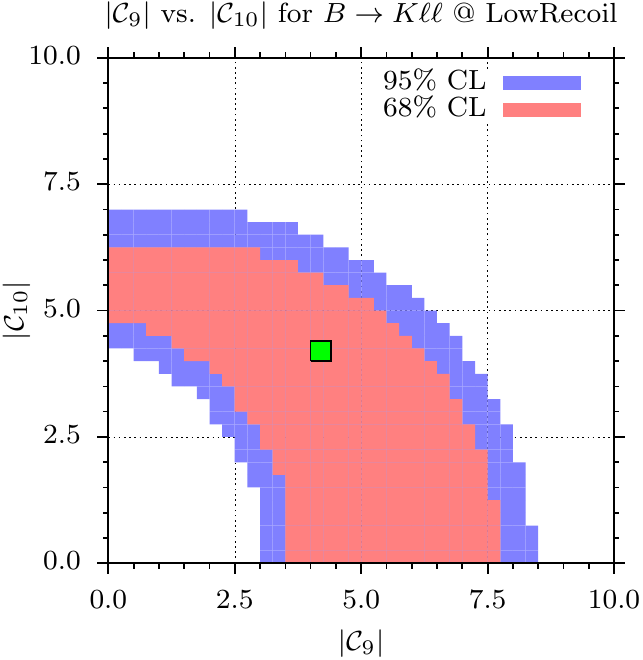}
  \includegraphics[width=.32\textwidth]{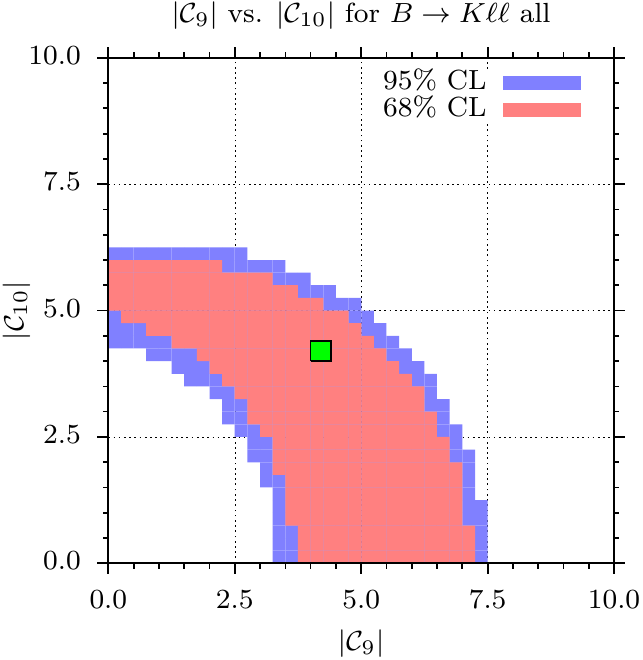}
  \caption{The constraints from $\bar{B}\to \bar{K} \ell^+ \ell^-$ decays on 
    $|\wilson{9}|$ and $|\wilson{10}|$ from the large-recoil region (left-hand plot), the low-recoil region (plot in the middle) and both (right-hand plot), for further notation see \reffig{constraints-c9-c10}.
    \label{fig:constraints-c9-c10-history}
   }
\end{figure}

\begin{figure}[t]
  \includegraphics[width=.49\textwidth]{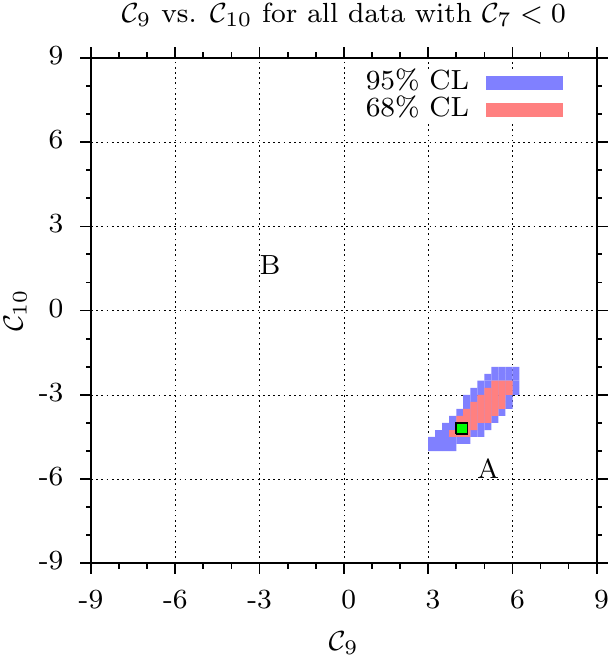}
  \includegraphics[width=.49\textwidth]{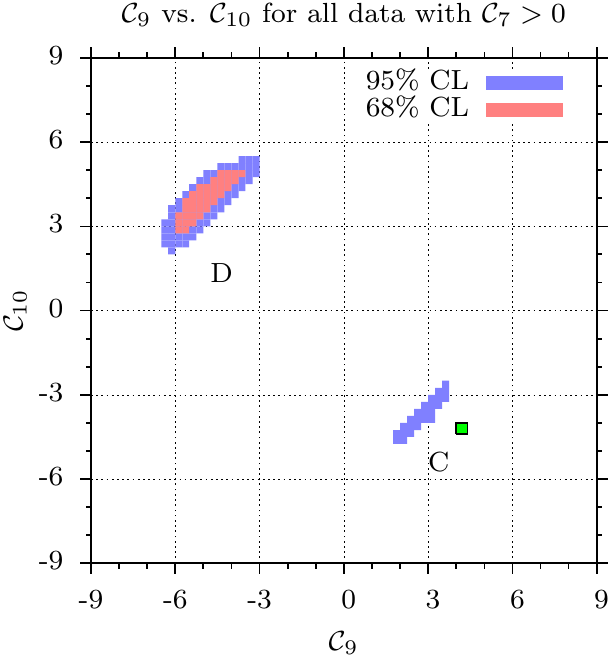}
  \caption{The constraints from all the currently available data on the real-valued Wilson coefficients $\wilson{9}$ and $\wilson{10}$ for SM-like sign $\wilson{7}<0$ (left-hand plot) and 
  $\wilson{7}>0$ (right-hand plot),
         for further notation see \reffig{constraints-c9-c10}.
    \label{fig:constraints-c9-c10-real}
   }
\end{figure}

\begin{figure}[t]
  \includegraphics[width=.8\textwidth]{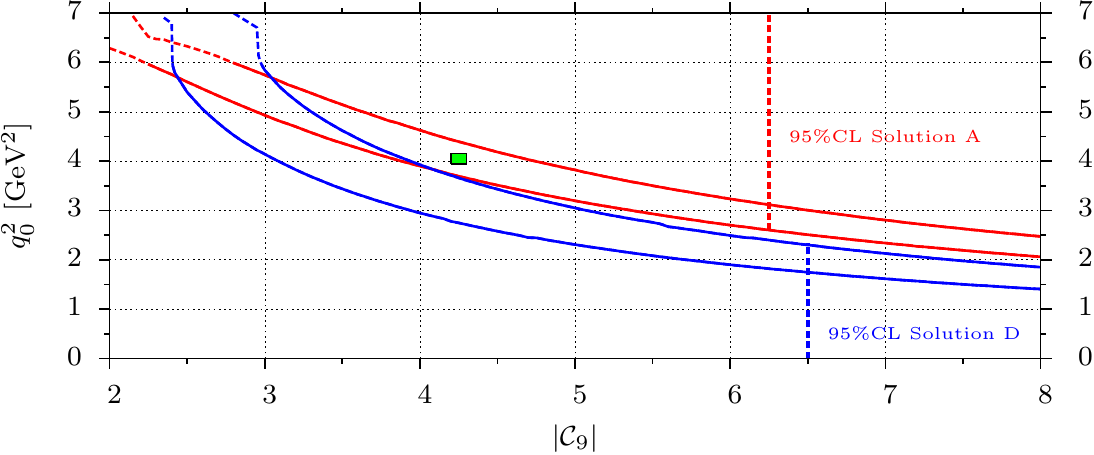}
\caption{The zero of the forward-backward asymmetry $q_0^2$
in $\bar{B}^0\to\bar{K}^{*0}\ell^+\ell^-$ decays at 1 $\sigma$ versus $|\wilson{9}|$ for
$\wilson{7}<0$  (solution A in \reffig {constraints-c9-c10-real}, red lines) and  $\wilson{7}>0$  (solution D  in \reffig {constraints-c9-c10-real}, blue lines) assuming no BSM CP-violation.
    The vertical lines denote
    the corresponding 95\% CL upper limits on $|\wilson{9}|$ from \reffig {constraints-c9-c10-real}.
    The green area marks the SM (for fixed $\mu=4.2 \GeV$).
    \label{fig:afb-zero-crossing}
  }
\end{figure}

\begin{table}[t]
\begin{tabular}{|l|lr|lr|} \hline
\multicolumn{1}{|c|}{Observable} & \multicolumn{2}{c|}{SM Prediction} &
\multicolumn{2}{c|}{Measurement}\\
\hline \hline
  $  \mathcal{B}(\bar{B}^0\to\bar{K}^0\mu^+\mu^-)_{1.0,6.0}$
& $\(1.59^{+0.59}_{-0.35}\) \times 10^{-7}$ & $^*$\cite{Bobeth:2007dw}
& $\(1.36^{+0.31}_{-0.29}\) \times 10^{-7}$ & $^\dagger$\cite{:2009zv}\\
  $  \mathcal{B}(B^-\to K^-\mu^+\mu^-)_{1.0,6.0}$
& $\(1.75^{+0.64}_{-0.38}\) \times 10^{-7}$ & $^*$\cite{Bobeth:2007dw}
& $\(1.41^{+0.29}_{-0.29}\) \times 10^{-7}$ & \cite{Aaltonen:2011qs}\\
\hline
  $\mathcal{B}(\bar{B}^0\to\bar{K}^0\mu^+\mu^-)_{14.18,16.00}$
& $\(0.34^{+0.18}_{-0.09}\) \times 10^{-7}$ &
& $\(0.38^{+0.21}_{-0.14}\) \times 10^{-7}$ & $^\dagger$\cite{:2009zv}\\
  $ \mathcal{B}(B^-\to K^-\mu^+\mu^-)_{14.18,16.00}$
& $\(0.37^{+0.20}_{-0.09}\) \times 10^{-7}$ &
& $\(0.53^{+0.13}_{-0.13}\) \times 10^{-7}$ & \cite{Aaltonen:2011qs}\\
\hline
  $  \mathcal{B}(\bar{B}^0\to\bar{K}^0\mu^+\mu^-)_{16.00,22.86}$
& $\(0.63^{+0.39}_{-0.18}\) \times 10^{-7}$ &
& $\(0.98^{+0.26}_{-0.24}\) \times 10^{-7}$ & $^\dagger$\cite{:2009zv}\\
  $  \mathcal{B}(B^-\to K^-\mu^+\mu^-)_{16.00,22.90}$
& $\(0.68^{+0.41}_{-0.19}\) \times 10^{-7}$ &
& $\(0.48^{+0.14}_{-0.14}\) \times 10^{-7}$ & \cite{Aaltonen:2011qs}\\
\hline
\end{tabular}
\caption{The $\bar B \to \bar K  \ell^+ \ell^-$ branching ratios and their respective measurements 
with systematic and statistical uncertainties added linearly
 by Belle \cite{:2009zv} and CDF \cite{Aaltonen:2011qs}  used to constrain $\wilson{7,9,10}$. 
All entries are understood as CP-averaged. The integration binning $q^2_{\rm min} \leq q^2 < q^2_{\rm max}$ in $\GeV^2$ is indicated as
$  \mathcal{B}_{q^2_{\rm min},q^2_{\rm max}}$. 
  $^*$Includes updates
 to the numerical input and uses a different parametrization of the form factors than \cite{Bobeth:2007dw}. $^\dagger$see footnote \ref{admixure}.}
  \label{tab:experiments}
\end{table}

We perform a model-independent analysis in the $\Delta B=1$
complex-valued Wilson coefficients $\wilson{k}=|\wilson{k}| \exp(i \phi_k)$, $k=7,9,10$.
We build on the analysis by  \cite{Bobeth:2010wg,Bobeth:2011gi} using the updated data from CDF \cite{Aaltonen:2011qs,Aaltonen:2011ja}, the first $\bar{B}\to\bar{K}^* \mu^+\mu^-$  data from LHCb \cite{LHCb:2011,Blake:2011ii}, and include $\bar{B}\to\bar{K}\ell^+\ell^-$ decays.
The corresponding experimental results for the latter branching ratios 
from Belle \cite{:2009zv} and CDF \cite{Aaltonen:2011qs}  are given in \reftab{experiments}, and
also shown  in \reffig{diff-br} (see also footnote \ref{admixure}).  The data by BaBar  \cite{Aubert:2006vb}
shown  as well in \reffig{diff-br}  are not used as none of the two bins are fully applicable to
low- or high-$q^2$ theory frameworks. We discard CDF's data
on ${\cal{B}}(\bar{B}^0\to\bar{K}^0\mu^+\mu^-)$  \cite{Aaltonen:2011qs}  as the uncertainties are very large.  First, binned measurements of $A_{\rm FB}^\ell$ for light leptons are available and 
are consistent with  $A_{\rm FB}^\ell=0$ \cite{:2009zv,Aaltonen:2011ja}. These data are  
not taken into account because within the context of the SM operator basis used in this analysis
 $A_{\rm FB}^\ell$ vanishes, see \refeq{AFB}.
Decays into ditaus have not been observed. The presently best limit
stems from BaBar ${\cal{B}}(B^+ \to K^+ \tau^+ \tau^-)< 3.3 \times 10^{-3}$ for $q^2 > 14.23 \GeV^2$ at 90\% CL \cite{Walsh:2011}.

We perform a
six-dimensional parameter scan over the magnitudes $|\wilson{7,9,10}|$ and weak phases 
$\phi_{7,9,10}$. The resolution in $|\wilson{9,10}|$  is 0.25, and $\pi/16$ in $\phi_{7,9,10}$. 
We vary $0.3 < |\wilson{7}| \leq 0.4$ with step-width 0.02.  Statistical and systematic experimental uncertainies are added linearly in the scan. Contrary to  \cite{Bobeth:2011gi} we do not symmetrize the uncertainties, however, we checked that with the current data the results 
exhibit insignificant differences only between using symmetrized and un-symmetrized uncertainties.
The constraints on $|\wilson{9}|$-$|\wilson{10}|$  from the joint $\bar B\to \bar K^*\ell^+\ell^-$, $\bar B\to \bar K\ell^+\ell^-$ and $\bar B\to X_s\ell^+\ell^-$ analysis are presented
in \reffig{constraints-c9-c10}.  The inner (outer) colored  areas are allowed at
68\% CL (95\% CL).  The inclusion of the decay $\bar{B}\to\bar{K}\ell^+\ell^-$ improves the scan, as can be seen from the overlaid  contours obtained without using $\bar{B}\to\bar{K}\ell^+\ell^-$ data. As  the latter enter the analysis only
via branching ratios,  they put constraints on $\rho_1$, roughly $\sim |\wilson{9}|^2+|\wilson{10}|^2$, and show very little sensitivity to CP phases \cite{DA-Wacker}.
In \reffig{constraints-c9-c10-history}  we show  the constraints from
$\bar B \to \bar K \ell^+ \ell^-$ decays alone in the large recoil region, the low recoil region and
for both kinematic regions, respectively. All these constraints are consistent with each other and the  agreement with the SM is  good.

{}From the full scan we obtain  the allowed ranges
\begin{align} \label{eq:c9bound}
  2.3 & \leq |\wilson{9}|  \leq 6.5 \,, &  (1.0 & \leq |\wilson{9}|  \leq 7.0)\,,\\
  2.3 & \leq |\wilson{10}| \leq 5.3 \,, &  (1.8 & \leq |\wilson{10}| \leq 5.5)
   \label{eq:c10bound}
\end{align}
at $68\%$ CL ($95\%$ CL), see \reffig{constraints-c9-c10}.
This implies that
the branching ratios ${\cal{B}}(\bar{B}_s\to\ell^+\ell^- )\propto |\wilson{10}|^2$ for $\ell=e,\mu,\tau$ can be enhanced by at most a factor of 1.7 at $95\%$ CL with respect to their SM value.

Furthermore, we obtain constraints for real-valued Wilson coefficients
$\wilson{9,10}$ by discarding from the complex-valued scan all data with $\phi_{7,9,10} \neq 0, \pi$. These constraints are shown in \reffig{constraints-c9-c10-real}. 
A previously existing region, labelled B, is now excluded, while region C is disfavored at 95\% CL.
With CP-violation neglected the upper bound on $|\wilson{9}|$  provides
 a lower limit on the zero crossing $q^2_0$ of
the forward-backward asymmetry  in $\bar{B}\to\bar{K}^*\ell^+\ell^-$
decays \cite{Hiller:2011sg}. We find $q^2_0 > 1.7\,\GeV^2$, which gets improved to
$q_0^2 > 2.6 \,\GeV^2$  assuming the sign of $\wilson{7}$ to be SM-like, that is, negative.
The zero as a function of  $|\wilson{9}|$ (for fixed $\mu=4.2 \GeV$) for
$\bar{B}^0\to\bar{K}^{*0}\mu^+\mu^-$ decays is shown in \reffig{afb-zero-crossing} for both sign combinations which allow for a zero,  that is, solution A and D of   \reffig{constraints-c9-c10-real}. There is no such limit on $q_0^2$ in the fully CP-violating case. The dashed curves denote
extrapolations to $q^2 \gtrsim 6 \GeV^2$ where QCDF may not be applicable.

In the SM we find the zero to be located at
\begin{align}
    q^{2,{\rm SM}}_0 = \(3.97\,{^{+0.03}_{-0.03}|}_{\rm
FF}\,{^{+0.09}_{-0.09}|}_{\rm
SL}\,{^{+0.29}_{-0.27}|}_{\rm SD}\)\,\rm
{GeV}^2
\end{align}
consistent with \cite{Beneke:2004dp,Ali:2006ew,Egede:2008uy}.
The SD uncertainty without the one from the scale $\mu$ is $^{+0.16}_{-0.05} \, \mbox{GeV}^2$.
The location of the zero is form factor-independent only at lowest order in 
$1/m_b$ \cite{Ali:1999mm,Beneke:2001at}.
The uncertainty denoted by FF stems from varying the LCSR inputs 
to the form factors used in the scan to describe $\bar B \to \bar K^* \ell^+ \ell^-$ decays by  \cite{Ball:2004rg}. 

One can ask about the implications for generic new physics from the $\Delta B=1$ BSM  operators
$\sum_{i} \frac{\tilde c _i}{\Lambda_{\rm NP}^2}  \widetilde O_i $, where
\begin{align}
\widetilde O_{9} =   \bar s  \gamma_\mu (1-\gamma_5)b \, \bar \mu \gamma^\mu  \mu \, ,~~~~~~
\widetilde O_{10}=  \bar s  \gamma_\mu (1-\gamma_5)b \, \bar \mu \gamma^\mu \gamma_5 \mu \, .
\end{align}
Assuming unsuppressed contributions, $|\tilde c_{9,10}| =1 $, the scale of new physics $\Lambda_{\rm NP} $ must be as high as
\begin{align}
\Lambda_{\rm NP}  & > 30 \, \mbox{TeV} ~(15 \, \mbox{TeV}) \, ~~~~~ \mbox{from $ \widetilde O_{9}$} , \\
\Lambda_{\rm NP}  & > 44 \, \mbox{TeV} ~(16 \, \mbox{TeV}) \, ~~~~~ \mbox{from $ \widetilde O_{10}$} .
\end{align}
On the other hand, with $\Lambda_{\rm NP}=1$ TeV
the coefficient of the higher dimensional operator $\tilde O_{10}$ needs a (flavor) suppression 
as strong as $|\tilde c_{10}| <  5 \times 10^{-4} \, (4 \times 10^{-3})$. The corresponding numbers for
 $\tilde O_{9}$ read $|\tilde c_{9}| <  1 \times 10^{-3} \, (5 \times 10^{-3})$.
The bounds are obtained  at 95\% CL from Eqs.~(\ref{eq:c9bound}) and (\ref{eq:c10bound}). For all cases
the first (second) number corresponds to constructive (destructive) interference with the SM.
The bounds from $\wilson{10}$ are stronger than those of $\wilson{9}$.

%
%
%--------+---------+---------+---------+---------+---------+---------+---------+
\section{Conclusion \label{sec:conclusion}}

With event samples of order several hundred analyzed \cite{:2009zv,Aaltonen:2011qs,Aubert:2006vb,LHCb:2011,Blake:2011ii} and beyond a  thousand from LHCb alone at the horizon \cite{Golutvin:2011}
the studies of  rare decays $\bar{B}\to\bar{K}^{(*)} \ell^+\ell^-$ 
begin to reach deep into  the space of $\Delta B=1$ weak operators in terms of
the  allowed ranges and correlations of the short-distance couplings.

Allowing for BSM CP-violation the current status is shown in \reffig{constraints-c9-c10}, updating the analysis of \cite{Bobeth:2011gi} for the moduli of the Wilson coefficients $\wilson{9}$ and $\wilson{10}$ by including the data from LHCb \cite{LHCb:2011}
and $\bar{B}\to\bar{K} \ell^+\ell^-$ decays.
For the latter we worked out predictions for the kinematic region of low hadronic recoil, proven already beneficial for
$\bar{B}\to\bar{K}^*\ell^+\ell^-$ decays \cite{Bobeth:2010wg}. The low recoil framework is equally applicable to other $B$-meson to pseudo-scalar decays, including $\bar B \to \pi \ell^+ \ell^-$, $\bar B_s \to \bar K \ell^+ \ell^-$
and $\bar B_s \to \eta^{(\prime)} \ell^+ \ell^-$.
We find that the SM is  presently in good agreement with the rare decay data.

Assuming none or negligible  CP-violation only
the current situation is shown in \reffig{constraints-c9-c10-real}.
The preferred confidence region is shared between disconnected islands, one including the SM. Only two, solution A  and D, allow for a forward-backward asymmetry zero in
$\bar{B}^0\to\bar{K}^{*0} \ell^+\ell^-$ decays. We predict  the location of the zero 
to be above $1.7 \GeV^2$. If a zero
is found, the solution C, which is already disfavored, is excluded. Vice versa, if no zero is found, solutions A, the SM, and D are excluded. To distinguish the solutions A and D
requires that the precision of the model-independent analysis reaches the level of being sensitive
to interference terms with SM dominated four-fermion operators.

The presence of the pseudo-scalar channel improves the statistical power of the analysis.
Even stronger advantages arise  
from correlations  among the rare decays: 
We find that at low recoil the $\bar{B}\to\bar{K} \ell^+\ell^-$ amplitude involves the same 
short-distance couplings present in 
$\bar{B}\to\bar{K}^*\ell^+\ell^-$ for light leptons ($\ell=e$ or $\mu$), and the corresponding CP-asymmetries in the rate, 
$a_{\rm CP}^{(1)}$ as in  \refeq{acp1}, are identical.
Moreover, the
angular distribution in $\bar{B}\to\bar{K}\ell^+\ell^-$  is an indicator for
operators beyond \refeq{Heff}. In this operator basis, 
$A_{\rm FB}(\bar{B}\to\bar{K}\ell^+\ell^-)$ vanishes and
$F_H^\ell$ receives only small  corrections of the order $m_\ell^2/m_B^2$, hence  vanishes for practical purposes for the light leptons as well.
These observables  will therefore be important null tests for effects from BSM operators, complementing the
search for $\bar B_s \to \mu^+ \mu^-$ decays and lepton non-universality
by comparing $\bar{B}\to\bar{K}\ell^+\ell^-$ observables into electrons with those into muons, e.g.,  \cite{Bobeth:2007dw}.

\medskip

{\it Note added:} During the completion of this work a systematic analysis of $\Delta B=1$ constraints including 
low- and high-$q^2$ data on $\bar{B}\to\bar{K}^*\ell^+\ell^-$  but without  $\bar{B}\to\bar{K} \ell^+\ell^-$ decays  appeared \cite{aps11}. The findings for the SM operator basis are
qualitatively consistent with \cite{Bobeth:2010wg,Bobeth:2011gi} and the present work.

%--------+---------+---------+---------+---------+---------+---------+---------+
\acknowledgments

We thank Frederik Beaujean for advice on multidimensional analyses, Hideki Miyake for uncovering a numerical instability
in the calculation of $\bar{B}\to\bar{K}\ell^+\ell^-$ observables in a prerelease
version of EOS and Mikihiko Nakao  for useful communications on Belle's 
$B \to K^{(*)} \ell^+ \ell^-$ measurements.
We are thankful to the technical team of the $\Phi$Do HPC cluster,
without which we could not have performed our scans.

%
%--------+---------+---------+---------+---------+---------+---------+---------+
\appendix

%
%
%--------+---------+---------+---------+---------+---------+---------+---------+
\section{The $B \to K$ form factors \label{app:FFs}}

\begin{figure}[t]
\begin{center}
\includegraphics[width=.9\textwidth]{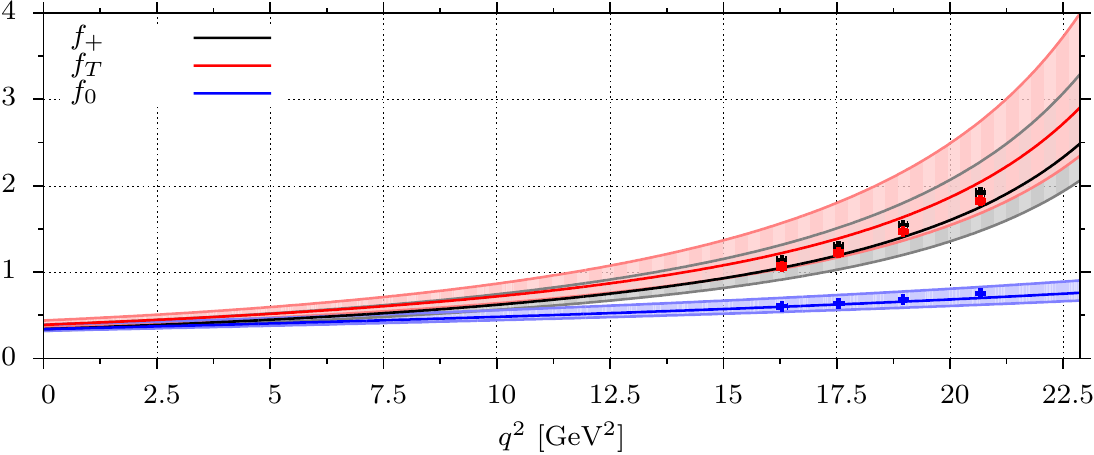}
\end{center}
\caption{
The  $\bar{B}\to\bar{K}$ form factors $f_+$, $f_T$ and $f_0$ by \cite{Khodjamirian:2010vf} extrapolated to the high-$q^2$ endpoint and preliminary lattice results  with statistical uncertainties only  (data points)  \cite{Liu:2011ra}. The shaded bands show the
  respective form factor uncertainties. \label{fig:form-factors}}
\end{figure}

\begin{figure}[t]
\includegraphics[width=.9\textwidth]{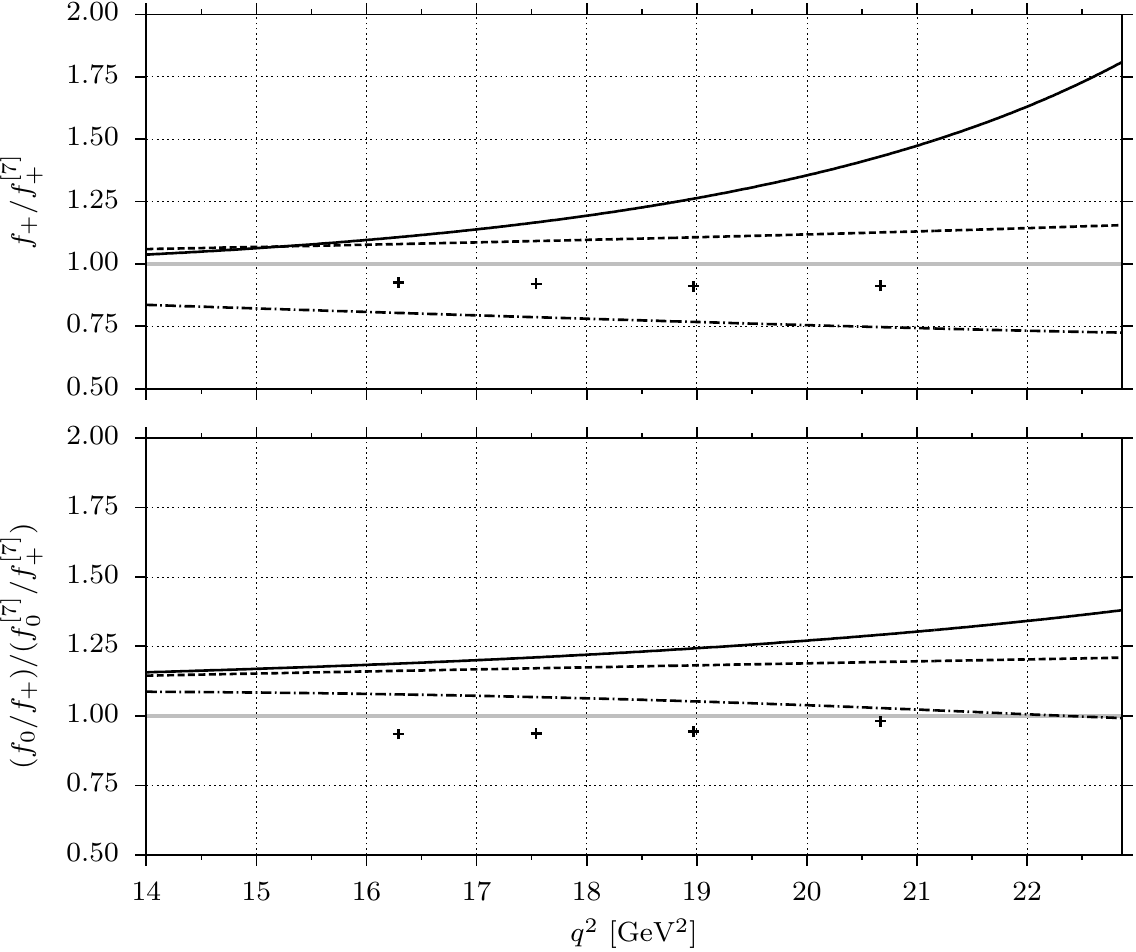}
\caption{Ratios of $f_+$  and $f_0/f_+$ from
extrapolated LCSR by \cite{Ball:2004ye} (solid lines), extrapolated LCSR with simplified
series expansion \cite{Bharucha:2010im} (dashed),
a relativistic quark model  \cite{Faessler:2002ut} (dash-dotted)
and unquenched lattice calculations  \cite{Liu:2011ra} (points),
over the corresponding extrapolated form factors  \cite{Khodjamirian:2010vf} used in this work.}
\label{fig:ffcomp-relative}
\end{figure}

\begin{figure}[t]
\includegraphics[width=.9\textwidth]{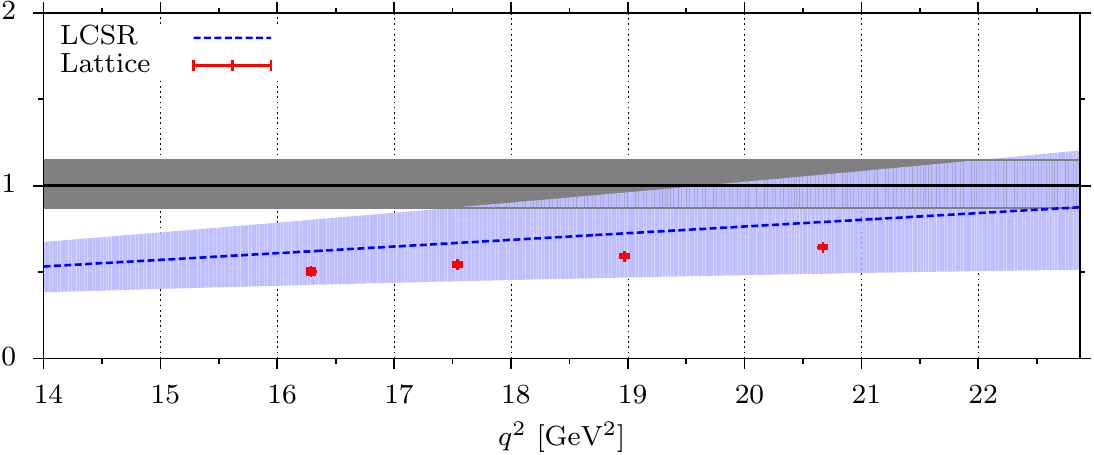}
\caption{The relation $R_T(q^2)$ given in \refeq{R_T} from extrapolations  of the LCSR form factors of \cite{Khodjamirian:2010vf} (blue band) and lattice \cite{Liu:2011ra} (red points). The grey band indicates the $\Lambda/m_b$-correction
to \refeq{IW-relation}, see text.
   \label{fig:ff-relations}}
\end{figure}

The $B \to K$ form factors $f_{+,T,0}$ are defined as usual
\begin{align} \label{eq:fT}
  \bra{\bar{K}(k)} \bar{s}\gamma^\mu b\ket{\bar{B}(p)}
    & = f_+(q^2) (p + k)^\mu + \[f_0(q^2) - f_+(q^2)\] \frac{m_B^2 - m_K^2}{q^2} q^\mu\,,
\\
  \bra{\bar{K}(k)} \bar{s}\sigma^{\mu\nu} b\ket{\bar{B}(p)}
    & = i \frac{f_T(q^2)}{m_B + m_K} \[(p + k)^\mu q^\nu - q^\mu (p + k)^\nu\]\,,
\end{align}
where $p$ ($k$) denotes the four-momentum of the $B$-meson (kaon) and $q = p - k$. 
The numerical analysis in this work is performed with the form factors from Ref.~\cite{Khodjamirian:2010vf} parameterized as, $i=+,T,0$
\begin{align} \label{eq:ffparam}
f_{i}(s)=\frac{f_{i}(0)}{1-s/m_{\text{res},i}^{2}} 
\{ 1+b_{1}^{i} \left( z(s)-z(0)+\frac{1}{2}\, ( z(s)^{2}-z(0)^{2} ) \right) \} , ~~~s=q^2 , \\
z(s)=\frac{\sqrt{\tau_{+}-s}-\sqrt{\tau_{+}-\tau_{0}}}{\sqrt{\tau_{+}-s}+\sqrt{\tau_{+}-\tau_{0}}}, ~~~\tau_{0}
=\sqrt{\tau_{+}}\left(\sqrt{\tau_{+}}-\sqrt{\tau_{+}-\tau_{-}}\right), ~~~
\tau_{\pm} =\left(m_{B}\pm m_{K}\right)^{2}  \, , \nonumber
\end{align}
and the input given in \reftab{ffinput}. The values $f_i(q^2=0)$ stem from LCSR calculations.
The form factors are shown in \reffig{form-factors}, extrapolated to high $q^2$.
Here, the agreement with preliminary lattice results by Liu {\it et al.} \cite{Liu:2011ra}, which are shown as well, is good.

\begin{table}[t]
\begin{tabular}{|c|c|c|c|}
\hline
Form Factor $f_{i}$ & Resonance &
\textbf{$f_{i}\left(0\right)$} & \textbf{$b_{1}^{i}$}\tabularnewline
\hline
\hline
$f_{+}$ & $m_{\text{res},+}=5.412\,\text{GeV}$ & $0.34_{-0.02}^{+0.05}$ &
$-2.1_{-1.6}^{+0.9}$\tabularnewline
$f_{0}$ & no pole & $0.34_{-0.02}^{+0.05}$ & $-4.3_{-0.9}^{+0.8}$\tabularnewline
$f_{T}$ & $m_{\text{res},T}=5.412\,\text{GeV}$ & $0.39_{-0.03}^{+0.05}$ &
$-2.2_{-2.0}^{+1.0}$\tabularnewline
\hline
\end{tabular}
\caption{Input to the $B \to K$ form factor parameterization \refeq{ffparam} from \cite{Khodjamirian:2010vf}.}
\label{tab:ffinput}
\end{table}

Form factors stemming from different methods and parameterizations relevant to
$\bar B \to \bar K \ell^+ \ell^-$ decays at low recoil are compared in \reffig{ffcomp-relative}. Here we show $f_+$ and the ratio $f_0/f_+$ entering $\bar B \to \bar K \tau^+ \tau^-$ decays divided by the corresponding (extrapolated) ones used in our analysis, from Ref.~\cite{Khodjamirian:2010vf}. 
Owing to a different $q^2$-shape the extrapolated LCSR ones
from  \cite{Ball:2004ye} (solid lines) grow much larger towards very low recoil,
and differ most strongly from the findings of \cite{Khodjamirian:2010vf} and lattice  
\cite{Liu:2011ra}. Both findings  for  $f_+$ in the relativistic quark model  \cite{Faessler:2002ut} and LCSR  from  \cite{Ball:2004ye} combined with a simplified
series expansion \cite{Bharucha:2010im}  exhibit a  shape similar to the ones from
\cite{Khodjamirian:2010vf}. The differences between
$f_+$ and $f_0/f_+$  from  \cite{Khodjamirian:2010vf} and \cite{Faessler:2002ut,Bharucha:2010im} are within $25\%$ and
$21\%$, respectively, which is within the uncertainties covered by \reftab{ffinput}.

The validity of the lowest order improved Isgur-Wise form factor
relation  \refeq{IW-relation} can be quantified  by looking at deviations of
\begin{align}
  R_T(q^2) & = \frac{q^2}{m_B(m_B + m_K)} \frac{f_T(q^2)}{f_+(q^2)}
  \label{eq:R_T}
\end{align}
from $\kappa$, given in \refeq{kappa}.
As can be inferred from
\reffig{ff-relations}, for both the LCSR extrapolation \cite{Khodjamirian:2010vf} and lattice results
\cite{Liu:2011ra}  the agreement is good near the kinematical endpoint. For smaller dilepton masses $R_T$ is smaller than $\kappa \simeq 1$.
The agreement improves somewhat if the kinematical prefactor on the right-hand side of \refeq{R_T}  is replaced by one.
Note that $R_T(q^2)$ obtained with the form factors   from  \cite{Ball:2004ye,Bharucha:2010im,Faessler:2002ut} behaves very similar to the one of \cite{Khodjamirian:2010vf}.
Including a positive-valued $1/m_b$ HQET form factor $\delta_+^{(0)}(q^2)$, as, for instance, from \cite{Grinstein:2002cz}, the ratio $f_T/f_+$ calculated from  \refeq{IW-relation-sl-terms} would increase further.
The form factors $\delta_\pm^{(0)}$ are defined in terms of the 
heavy $b$-quark field $h_v$ as  \cite{Grinstein:2002cz}
\begin{align}
  \bra{\bar{K}(k)} \bar{s}\, i \!\stackrel{\leftarrow}{D}_{\mu} \! h_v \ket{\bar{B}(p)}
    & = \delta_+^{(0)}(q^2) (p + k)_\mu + \delta_-^{(0)}(q^2) q_\mu \, .
\end{align}
Lattice studies  for $B \to \pi$  \cite{Dalgic:2006dt} suggest that they are not larger than the estimate by dimensional analysis.

We recall that the impact of $f_T$ on  the $\bar B \to \bar K \ell^+ \ell^-$ decay observables at low recoil is subleading. This hampers on one side the extraction of $f_T$ from data, on the other side
reduces the theoretical uncertainties. Ultimately it is desirable to know  all form factors  more precisely from lattice QCD, or other means.

%
%
%--------+---------+---------+---------+---------+---------+---------+---------+

\end{document}